  \documentclass[aps,onecolumn]{revtex4-1}
\usepackage{graphicx,amsmath}\usepackage{color}
\draft 

\begin{document}


\title{Quantum carpets:
A probe to identify wave packet fractional revivals
} 



\author{Iqra Yousaf}

\affiliation{School of Natural Sciences, National University of Sciences and
             Technology, Islamabad, Pakistan}
\author{Shahid Iqbal}
\affiliation{School of Natural Sciences, National University of Sciences and
             Technology, Islamabad, Pakistan}
\email{sic80@hotmail.com, siqbal@sns.nust.edu.pk}


\date{\today}

\begin{abstract}

Quantum carpets$-$ in position and momentum space$-$ woven by the self-interference of de Broglie wave of an atom or an electron, trapped in an infinitely deep potential well, are explained. The recurrence of self-similar structures in designs of these carpets mimics the phenomena of quantum revivals and fractional revivals. We identify fractional revivals of various order by means of these space-time and momentum-time interference patterns.
\end{abstract}

\maketitle 


\section{Introduction}
Quantum world manifests a lot of phenomena which do not have analogous description in classical world$-$ nonclassical phenomena.
In general bounded Hamiltonian systems, quantum dynamics of a localized wave packet (WP) exhibits nonclassical features, for instance, quantum revivals and fractional revivals and formation of quantum carpets. 
An initially well-localized wave packet comprising many energy eigenstates of
a system follows classical trajectory in its short time evolution and displays reconstruction after a classical period.
Afterwards it spreads during its long time evolution, only to reverse the spreading and reshape after a certain time called the revival
time. Originally observed in highly excited
electrons in atoms \cite{jpb1,jpb2,jpb3}, the phenomenon has been
studied in a large variety of scenarios, such as, Gaussian WPs in exactly solvable models systems \cite{jpb17},
temporal evolution of coherent states \cite{gcs1,gcs2,gcs3}, periodically driven systems \cite{si2006},
WP dynamics even at an attosecond time scale  \cite{jpb9}, revivals in coherent photon
fields  \cite{jpb9a} or photon lattices \cite{jpb10}, or recently,
the propagation of WPs in graphene under magnetic fields \cite{jpb11,jpb13} and the use of WPs to detect extremely long coherence
times  \cite{jpb14}. Quantum revivals have also been widely studied over
the years in a more mathematical context, with emphasis on
obtaining accurate analytical expressions  \cite{jpb17,jpb15,jpb16,jpb18}. 

The multiple interferences of the WP components lead
also to the so-called fractional revivals at divisors of the revival
time, which have been predicted  \cite{jpb19,jpb20} and observed in Rydberg WPs  \cite{jpb21} or photon bouncing balls  \cite{jpb22}. Fractional revivals can retrieve information about the system even if it decays before the first revival  \cite{jpb23}. 
They can be used for mapping the quantum phase of a molecular nuclear WP in two-dimensional spectroscopy  \cite{jpb24}, or even to factorize prime numbers  \cite{jpb25}. It has also been shown that information entropy
in position and momentum spaces can reveal the existence of fractional revivals  \cite{jpb26}.

In this article the dynamical recurrences$-$ quantum revivals and fractional revivals$-$ of a Gaussian wave are explored by means of temporal evolution of probability density both in position and momentum spaces. The space-time and momentum-time evolution of the wave packet lead to the formation of highly structured patterns known as quantum carpets. The quantum carpets woven by position space wave packets have been discussed to explain the maximum and minimum probability density profiles \cite{qc1,qc2,qc3} and have been used to observe decoherence effects \cite{qc4} and Talbot effect with matter waves \cite{qc5}. However, quantum carpets woven by momentum space wave packets have only been discussed in few contexts \cite{qc6}. Here we show that the designs of quantum carpets reflect self-similar structures after regular intervals of space-time and momentum-time which leads us to identify dynamical fractional revivals. For our analysis we present a large gallery of quantum carpets and analyze the structure of fractional revivals.

The paper is organized as follows. 

In Sec. (\ref{model}), we discuss our model system and derive analytic expressions necessary for computing our numerical results. Moreover, we study the structure of quantum revivals and fractional revivals by means of autocorrelation function and present the related simulation results. In Sec. (\ref{qcxp}), we present a gallery a quantum carpets woven in position and momentum spaces and explore the structure of quantum revivals and fractional revivals by means of these carpets. We conclude our work in Sec. (\ref{sum}).
%

\section{Gaussian wave packet in infinitely deep well} \label{model}

The Hamiltonian of a particle of mass $m$ trapped inside infinitely deep quantum well of length $L$ is given as
\begin{equation}\label{H}
\hat{H}=\frac{\hat{P}}{2m}+V(x),
\end{equation} 
where the potential $V(x)=0$ if $0< x< L$ and $V(x)=\infty$ otherwise. 
The energy eigenstates and energy eigenvalues corresponding to the time-independent Schr\"{o}dinger equation are, respectively, given as
\begin{equation}\label{En}
u_{n}(x)=
\begin{cases}
    \sqrt{\frac{2}{L}}\sin{\frac{n\pi x}{L}}    & \text{if } 0< x< L,\\
    0     & \text{otherwise};\\
\end{cases} {\ \ \ \ \ \ } \text{and}{\ \ \ \ \ \ } E_{n}=\frac{n^{2}\pi^{2}\hbar^{2}}{2mL^{2}}; {\ \ \ \ \ \ }n=1,2,3...
\end{equation}
The time-dependent wave function for a localized wave packet is expanded as a superposition
of energy eigenstates $u_{n}(x)$ as
\begin{equation}\label{psixt}
\psi(x,t)=\sum_{n=0}^{\infty} c_{n}u_{n}(x)e^{-iE_{n}t/\hbar}.
\end{equation}
The expansion coefficient $c_{n}$ is calculated as
\begin{equation}\label{cn}
c_{n}=\int_{0}^{L}u_{n}(x)\psi(x,0)dx,
\end{equation}
where $\psi(x,0)$ is initial wave packet prepared at time $t=0$.
The recurrences of the wave packet at different time scales, classical periodicity, quantum revivals and fractional revivals, are conventionally identified by analyzing autocorrelation function \cite{jpb17} defined as
\begin{eqnarray}\label{At}
A(t)&=&\langle\psi(t)\vert\psi(0)\rangle=\int_{-\infty}^{\infty}\psi^{*}(x,t)\psi(x,0)dx,\nonumber\\
    &=&\sum_{n=0}^{\infty}\vert c_{n}\vert^{2}e^{iE_{n}t},
\end{eqnarray}
which is an overlap between the time-evolving and the initial wave packets. The modulus squared of the autocorrelation $\vert A(t)\vert^{2}$ takes a value between zero and one, such that for a complete overlap it is one and  for complete dephasing it is zero. 
For a wave packet which is narrowly peaked around a central value of quantum number $n_{0}$ with spread $\Delta n$, such that $n_{0}\gg\Delta n\gg1$, we can expand energy eigenvalues, given in Eq. (\ref{En}), by Taylor's expansion around $n_{0}$ as \cite{jpb17}
\begin{equation}\label{Enexp}
  E_{n}= E_{n_{0}}+E^{(1)}_{n_{0}}(n-n_{0})+\frac{E^{(2)}_{n_{0}}}{2!}(n-n_{0})^{2}
\end{equation}
where, $E^{(i)}_{n_{0}}=\frac{d^{(i)}E_{n}}{dn^{(i)}}\Big|_{n=n_{0}}$, is the $ith$ derivative of $E_{n}$ with respect to $n$ at $n=n_{0}$. The substitution of expansion (\ref{Enexp}) in to Eq. (\ref{At}) leads to identify the periodicities of the wave packet at different time scales. The first term in the expansion only produces a universal phase and does not contribute to the dynamics of the wave packet. However, second and third terms, respectively, define classical period, $T_{cl}$, and quantum revival time, $T_{rev}$, given as
\begin{figure}
\centering
\includegraphics[width=0.45\textwidth]{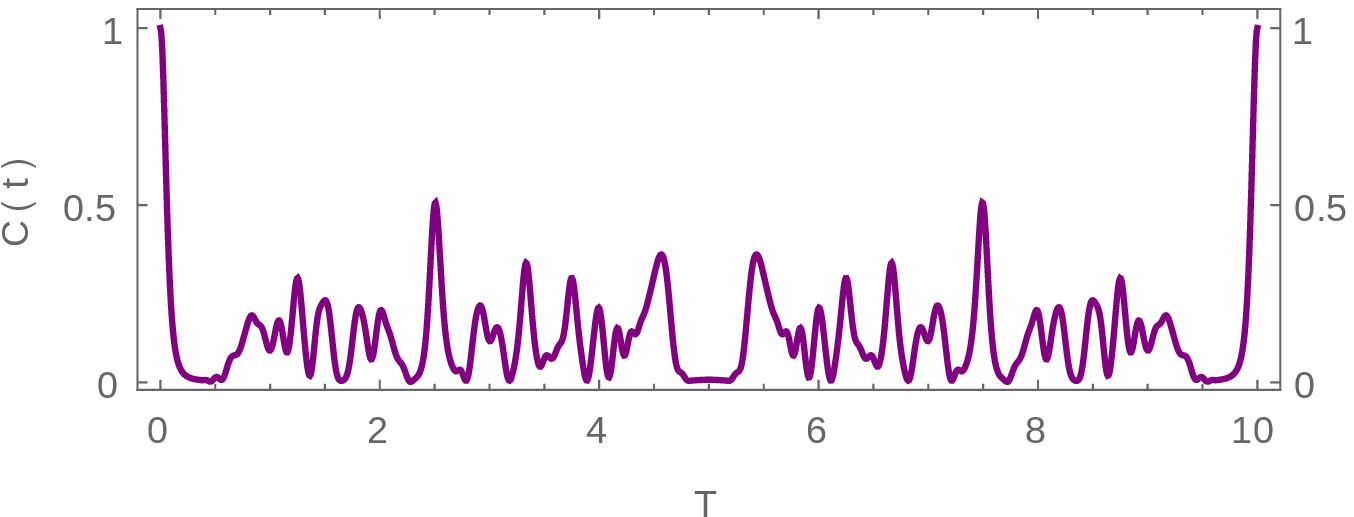}
\includegraphics[width=0.45\textwidth]{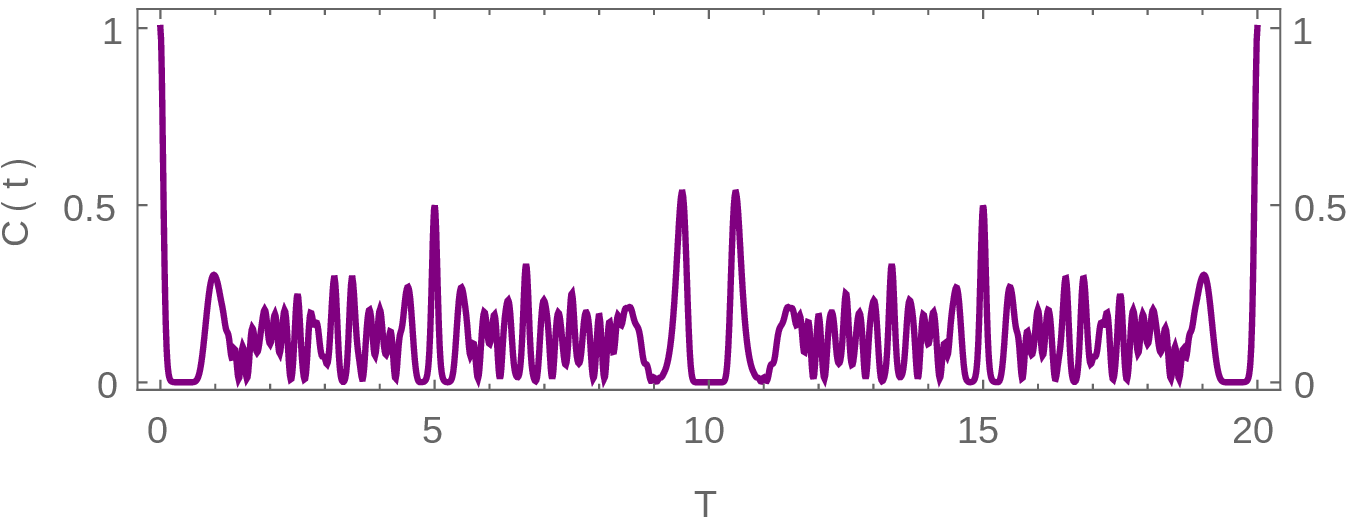}
\includegraphics[width=0.45\textwidth]{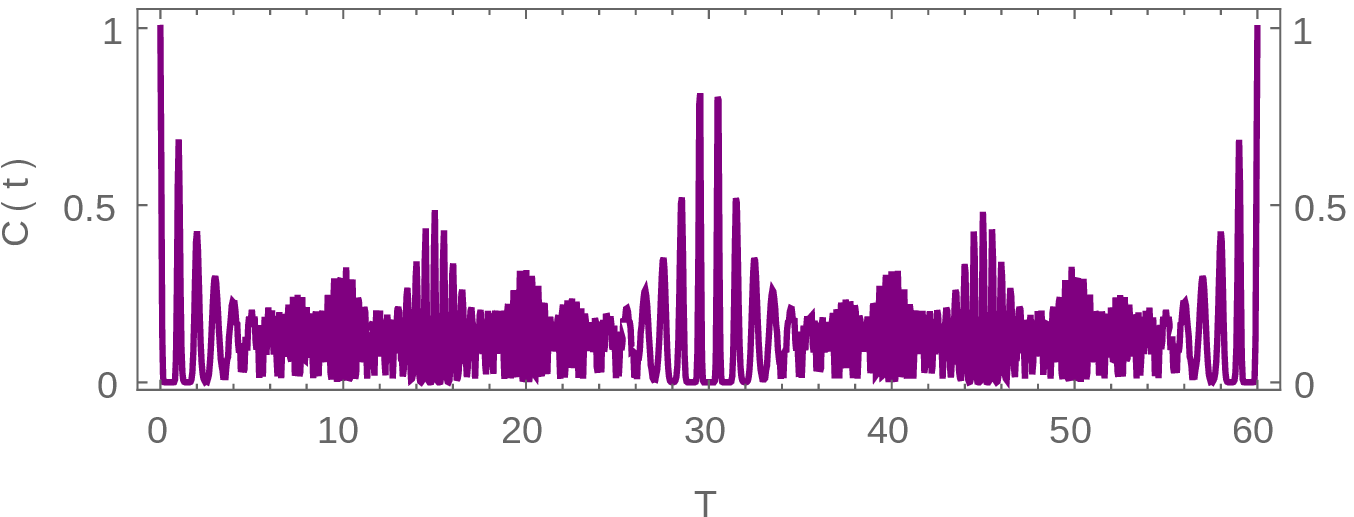}
\includegraphics[width=0.45\textwidth]{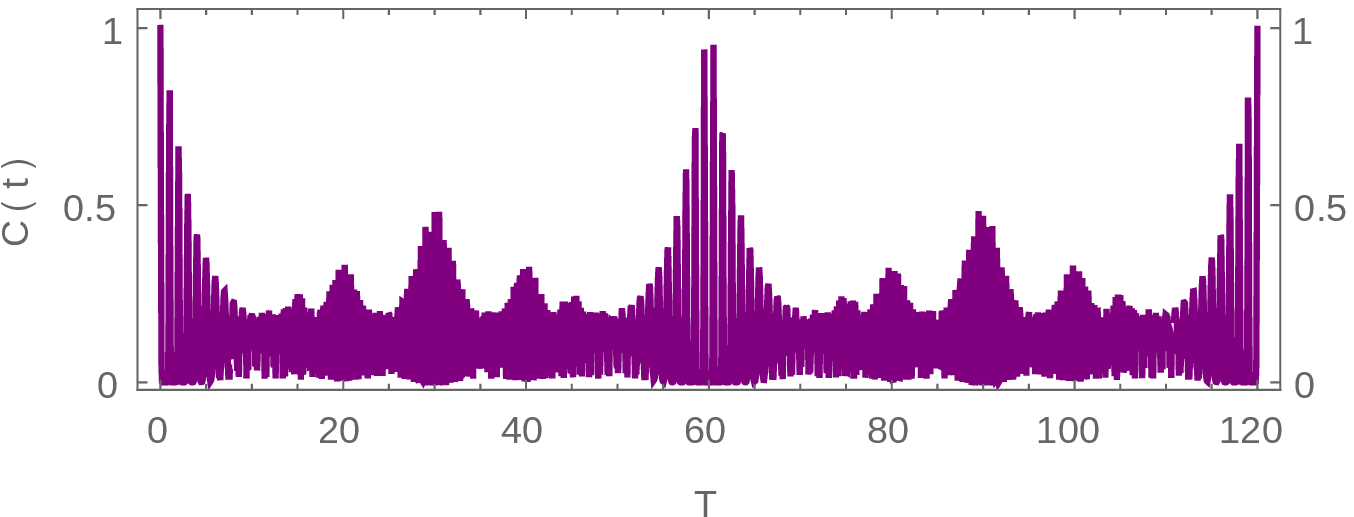}
\includegraphics[width=0.45\textwidth]{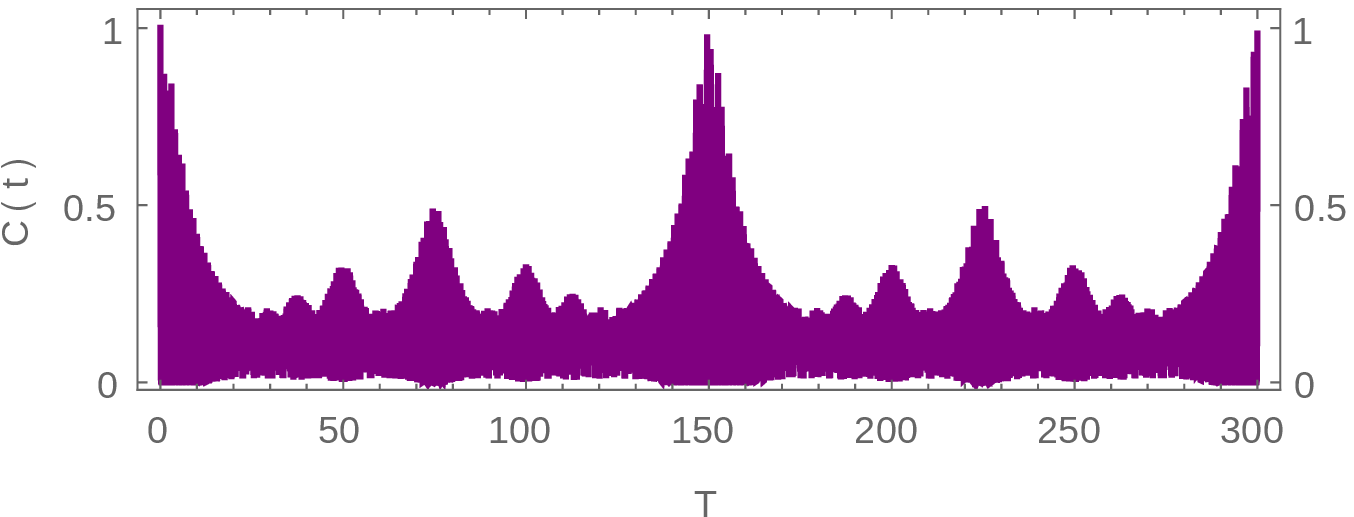}
\includegraphics[width=0.45\textwidth]{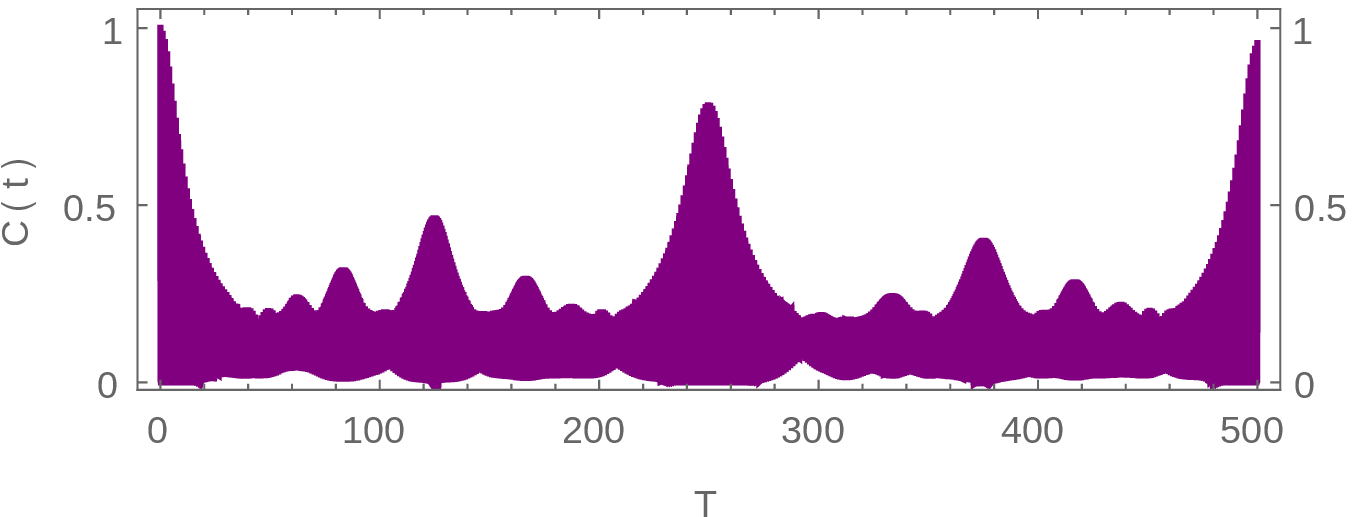}
\caption{Time evolution of $C(t)=\vert A(t)\vert^{2}$ for an initial Gaussian wave packer of width $\sigma=0.1$ centered at the middle of the well $x_{0}=L/2$ for different values of initial momentum $p_{0}=5\pi,10\pi,30\pi,60\pi,150\pi,250\pi$ from left-top to bottom-right, respectively. The $C(t)=\vert A(t)\vert^{2}$ is plotted for one revival time $T_{rev}$, where the horizontal axis is rescaled by classical period such that $T=t/T_{cl}$.}  \label{autofull}
\end{figure} 
\begin{equation}\label{tc}
 T_{cl}=\frac{2\pi\hbar}{|E_{n_{0}}^{(1)}|}, {\ \ \ \ \ \ \ \ \ \ \ } T_{rev}=\frac{2\pi\hbar}{|E_{n_{0}}^{(2)}/2!|},
\end{equation}
such that $T_{cl}<T_{rev}$. Using the value of $E_{n}$ from Eq.(\ref{En}), in Eq. (\ref{tc}) and taking the derivatives, the values of classical period $T_{cl}=2mL^{2}/ n_{0}\hbar\pi$ and quantum revival time $T_{rev}=4mL^{2}/ \hbar\pi$. 

In what follows, we consider an initial Gaussian wave packet
\begin{equation} \label{gaussian}
\psi(x,0)=\frac{1}{(\sigma \pi^{2})^{1/4}}e^{-(x-x_{0})^{2}/2\sigma^{2}}e^{ip_{0}x/\hbar},
\end{equation} 
centered at a position $x_{0}$, with a
width $\sigma$ and a mean momentum $p_{0}$. In this case, the expansion coefficients
$c_{n}$, given in Eq. (\ref{cn}), can be approximated with sufficiently high accuracy \cite{jpb17} which is given as
\begin{equation}\label{cn2}
c_{n}=\sqrt{\frac{4\sigma\pi}{L\sqrt{\pi}}}\frac{e^{-ip_{0}x_{0}/\hbar}}{2i}(e^{i n\pi x_{0}/L}e^{-\sigma^{2}(p_{0}+n\pi\hbar/L)^{2}/2\hbar^{2}}-e^{-i n\pi x_{0}/L}e^{-\sigma^{2}(p_{0}-n\pi\hbar/L)^{2}/2\hbar^{2}}),
\end{equation} 
where the region of integration is extended from $[0,L]$ to the whole real axis while obtaining the above result. For the sake of numerical computations, we will take $m=L=\hbar=1$. Henceforth, we will consider Gaussian wave packet, given in Eq. (\ref{gaussian}), with width $\sigma=0.1$ centered at $x_{0}=L/2=0.5$ throughout our ongoing analysis.

In Fig. (\ref{autofull}) modulus square of the autocorrelation function $\vert A(t)\vert^{2}=C(t)$ for the Gaussian wave packet 
with different values of initial momentum $p_{0}$ are plotted for one revival time $T_{rev}$, where the horizontal axis is rescaled by classical period such that $T=t/T_{cl}$. In this Fig. plots are shown for $p_{0}=5\pi,10\pi,30\pi,60\pi,150\pi,250\pi$, respectively, from left-top to bottom-right. It is obvious from the plots that the quantum wave packet revival occurs after (smaller) larger number of classical periods as the value of $p_{0}$ is (decreased) increased. This is due to the fact that $T_{rev}/T_{cl}=2n_{0}$ and $p_{n}=n\hbar\pi/L$ such that $p_{0}=n_{0}\hbar\pi/L$. For a wave packet with initial momentum $p_{0}=n\pi\hbar/L$, quantum revival occur after $2n$ (since $\hbar\L=1$)classical periods. 

In addition to the complete revival of the wave packet, fractional revivals of various order appear at times $t=T_{rev}p/q$, where $p,q$ are mutually prime numbers. It is obvious from Fig. (\ref{autofull}) that a complete mirror symmetry exists in the structure of fractional revivals about $t=T_{rev}/2$. Therefore, in order to analyze the fractional revivals we consider a time period equal to $T_{rev}/2$. 
\section{Quantum carpets in position and momentum spaces}\label{qcxp}
The dynamical behavior of a wave packet can, alternatively, be characterized by means of time evolution of probability density either in position space $\rho(x,t)=\vert\psi(x,t)\vert^{2}$ or in momentum space $\gamma(p,t)= \vert\phi(p,t)\vert^{2}$. The temporal evolution of either of these probabilities, for an initially well localized wave packet, undergoes a series of constructive and destructive interferences which leads to the formation of regular interference patterns known as quantum carpets. Here we present a large gallery of quantum carpets and analyze the structure of fractional revivals by means of these carpets.
\begin{figure}
\centering
\includegraphics[width=0.26\textwidth]{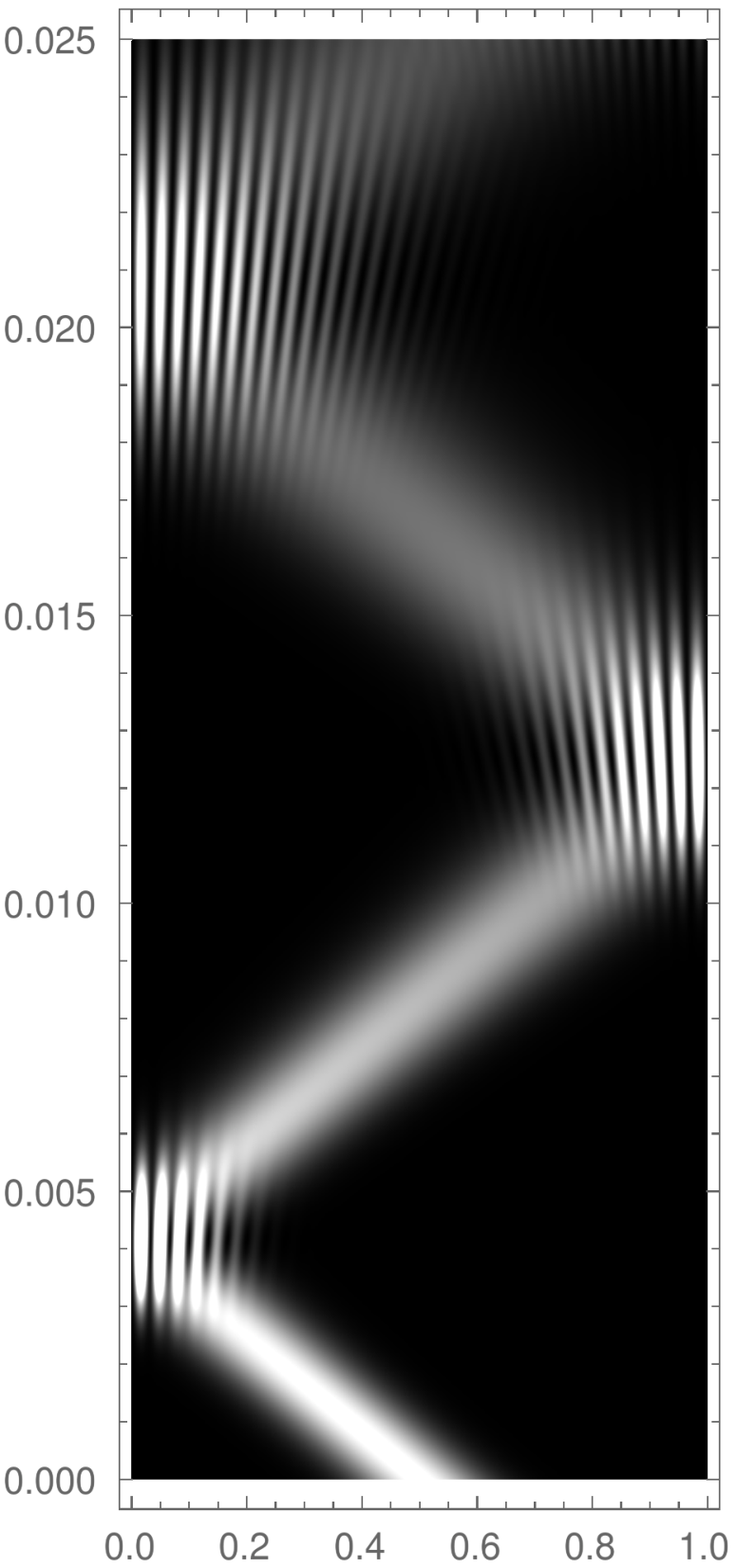}
\includegraphics[width=0.255\textwidth]{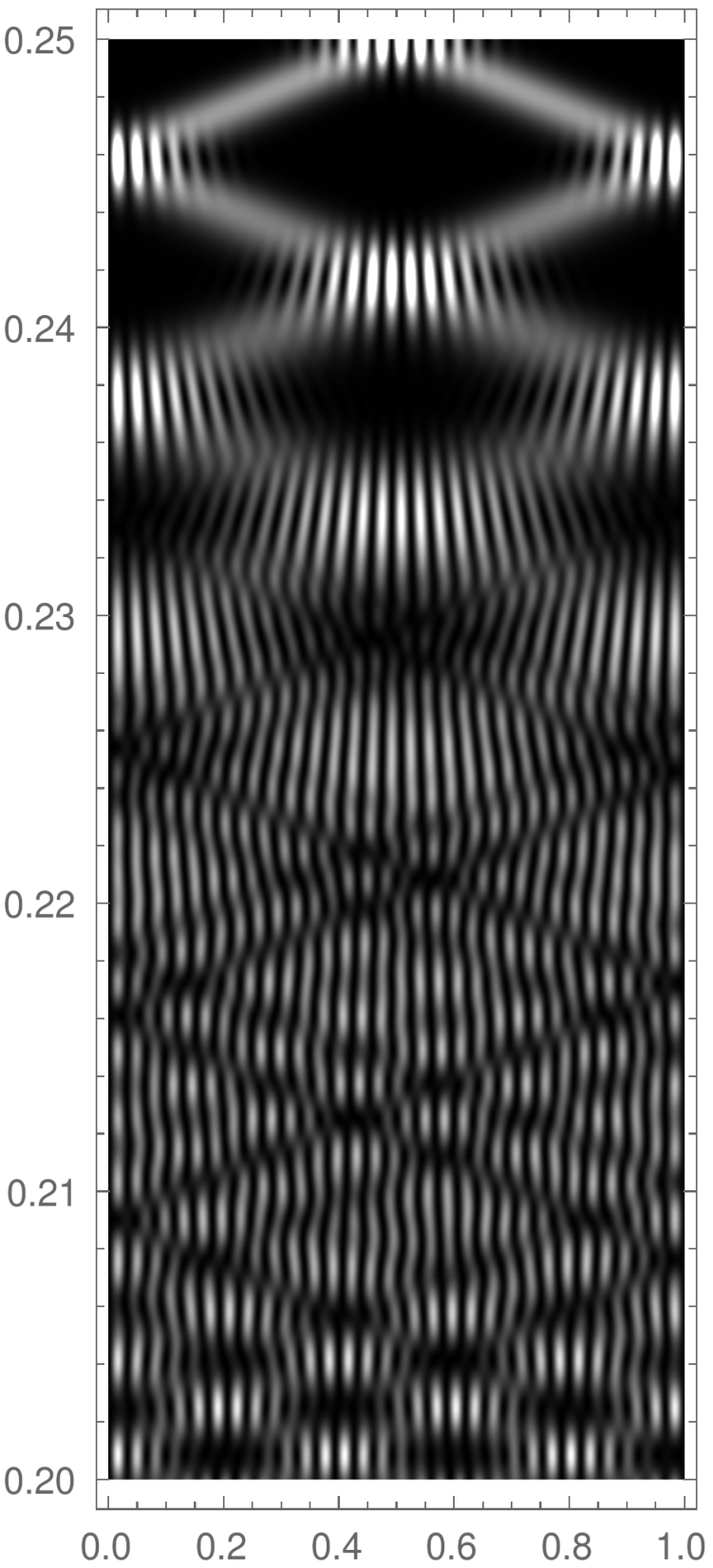}
\includegraphics[width=0.25\textwidth]{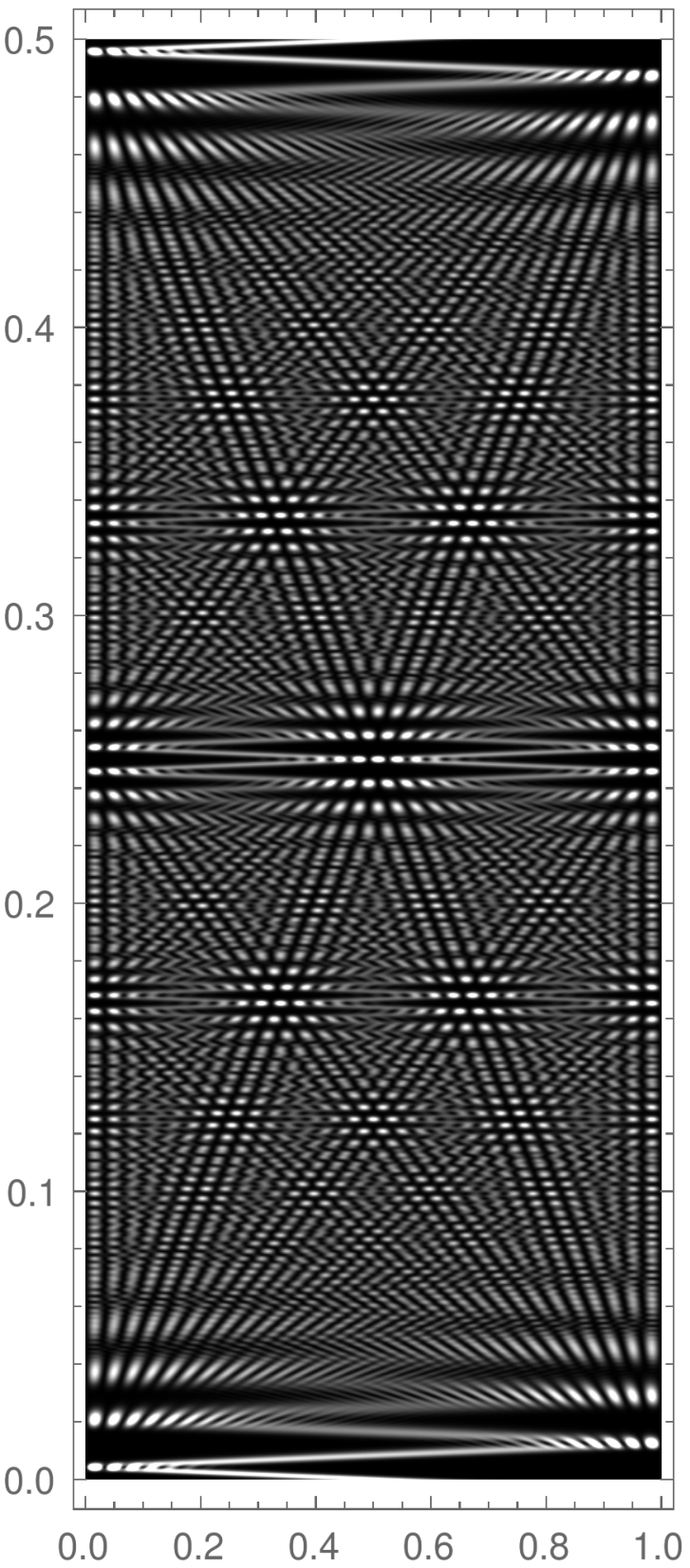}
\caption{Time evolution of position space probability density $\rho(x,t)=\vert\psi(x,t)\vert^{2}$ for an initial Gaussian wave packer with momentum $p_{0}=30\pi$ indicating (left to right) classical-like short time evolution which experiences a collapse and rebuilds its fractional copies of different order.} \label{qcxhalf}
\end{figure}
\begin{figure}
\centering
\includegraphics[width=0.25\textwidth]{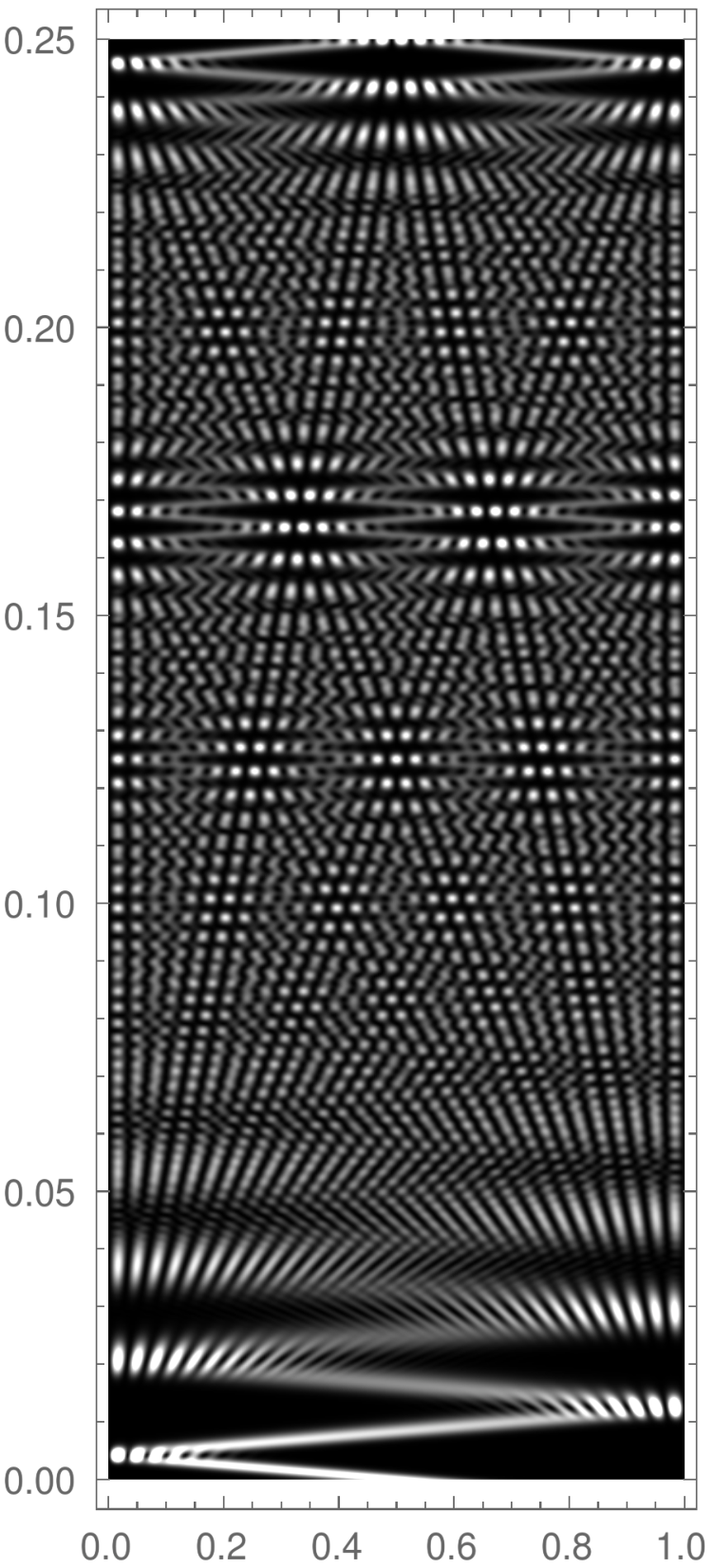}
\includegraphics[width=0.25\textwidth]{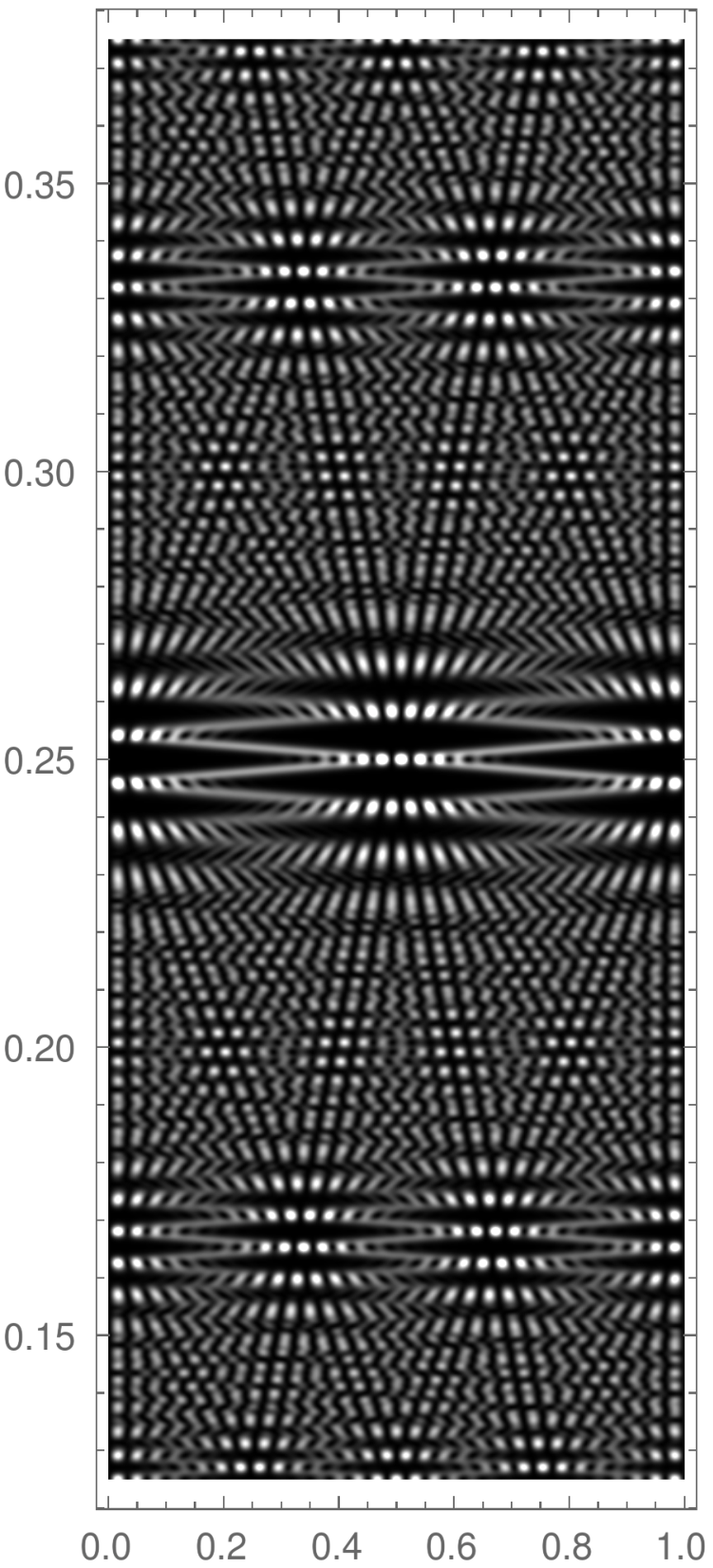}
\includegraphics[width=0.25\textwidth]{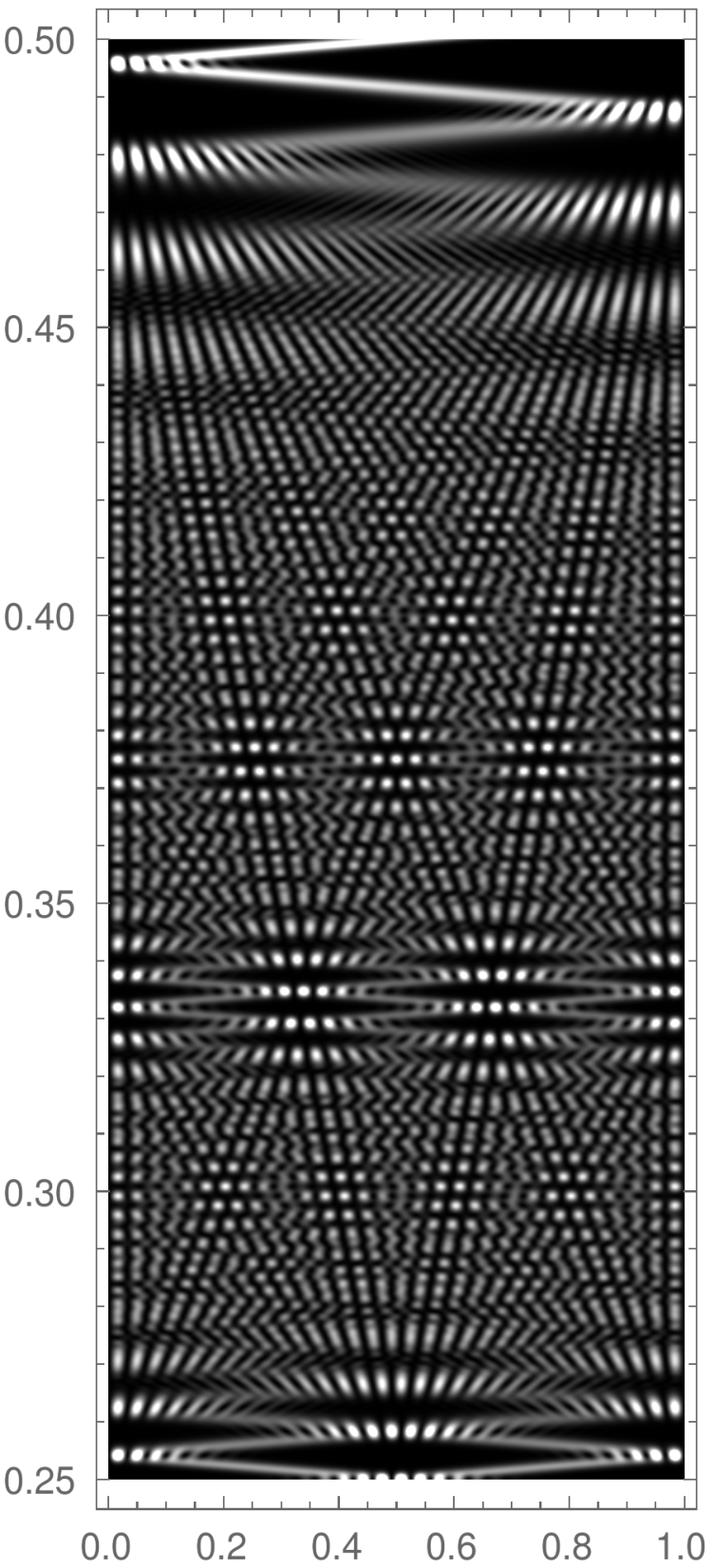}
\caption{Time evolution of $\rho(x,t)=\vert\psi(x,t)\vert^{2}$ for the same wave packer as in Figure (\ref{qcxhalf}) but fractional revivals of different order are highlighted.} \label{qcxfrac}
\end{figure}
The time evolution of probability density in position space can be calculated using time-dependent wave packet, given in Eq.(\ref{psixt}), as
\begin{equation}\label{rhoxt}
\rho(x,t)=\vert\psi(x,t)\vert^{2}=\sum_{n,m=0}^{\infty} c_{n}c_{m}^{*}u_{n}(x)u_{m}^{*}(x)e^{-i(E_{n}-E_{m})t/\hbar}.
\end{equation}
In order to obtain an analogous expression for time-dependent probability density in momentum space, we first calculate time-dependent
momentum-space wave packet by Fourier transform of $\psi(x,t)$ which is given as
\begin{equation}\label{phipt}
\phi(p,t)=\sum_{n=0}^{\infty} c_{n}\varphi_{n}(p)e^{-iE_{n}t/\hbar},
\end{equation}
where, $\varphi_{n}(p)$ is the momentum space eigenfunction obtained by the Fourier transform of $u_{n}(x)$ as
\begin{equation}\label{phipf}
  \varphi_{n}(p)=\frac{1}{\sqrt{2\pi\hbar}}\int_{-\infty}^{\infty}u_{n}(x)e^{-\frac{ipx}{\hbar}}dx.
\end{equation}
Using the position-space energy eigenfunctions $u_{n}(x)$, given in Eq.~(\ref{En}), and solving the Fourier integral, given in Eq.~(\ref{phipf}), the momentum-space eigenfunctions $ \varphi_{n}(p)$ are obtained \cite{jpb17} as
\begin{equation}\label{phip}
  \varphi_{n}(p)=\sqrt{\frac{\hbar}{\pi L}}\frac{p_{n}}{p^{2}-p_{n}^{2}}\left[(-1)^{n}e^{ipL/\hbar}-1\right],
\end{equation}
where $p_{n}=n\hbar\pi/L$. The time-dependent probability density in momentum space is now given as
\begin{equation}\label{gammapt}
\gamma(p,t)=\vert\phi(p,t)\vert^{2}=\sum_{n,m=0}^{\infty} c_{n}c_{m}^{*}\varphi_{n}(p)\varphi_{m}^{*}(p)e^{-i(E_{n}-E_{m})t/\hbar}.
\end{equation}
\begin{figure}
\centering
\includegraphics[width=0.25\textwidth]{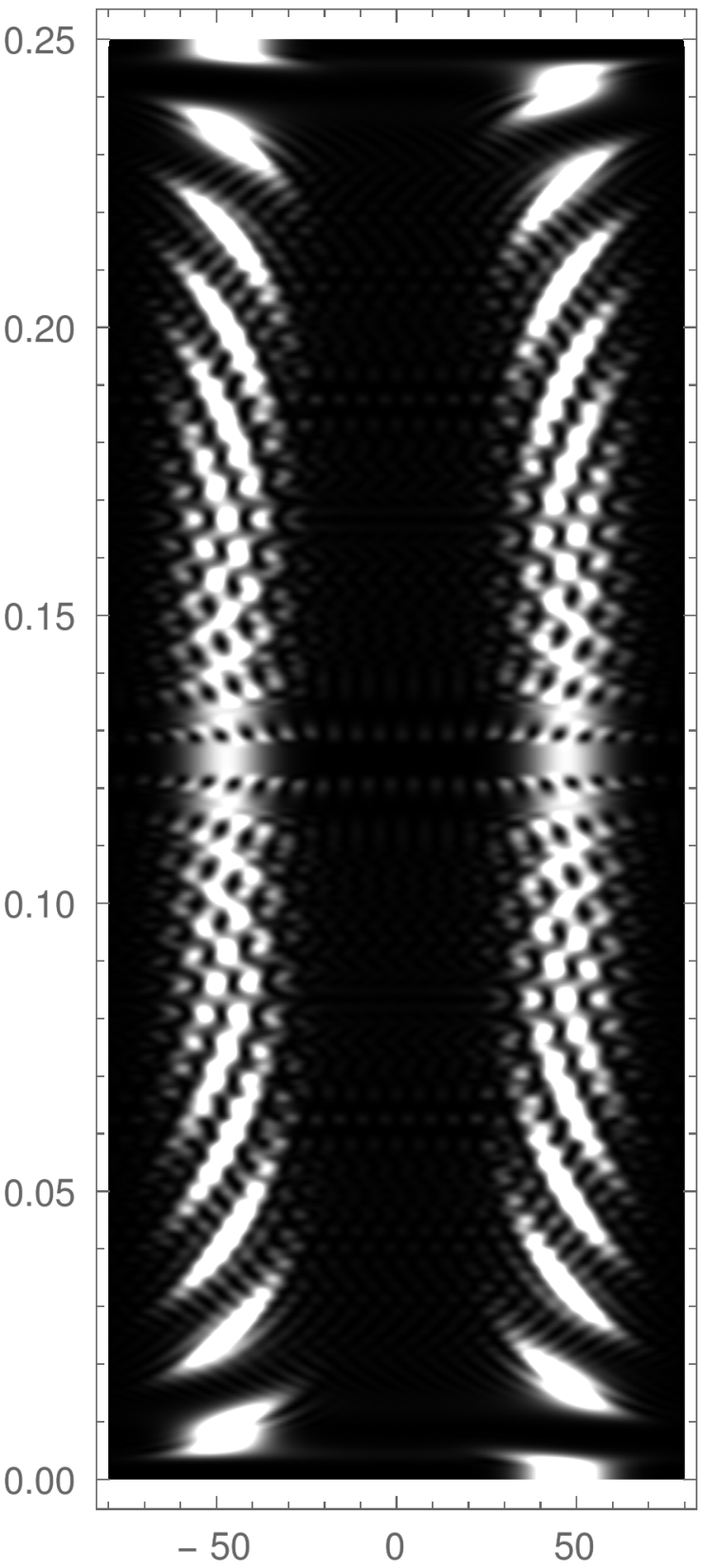}
\includegraphics[width=0.25\textwidth]{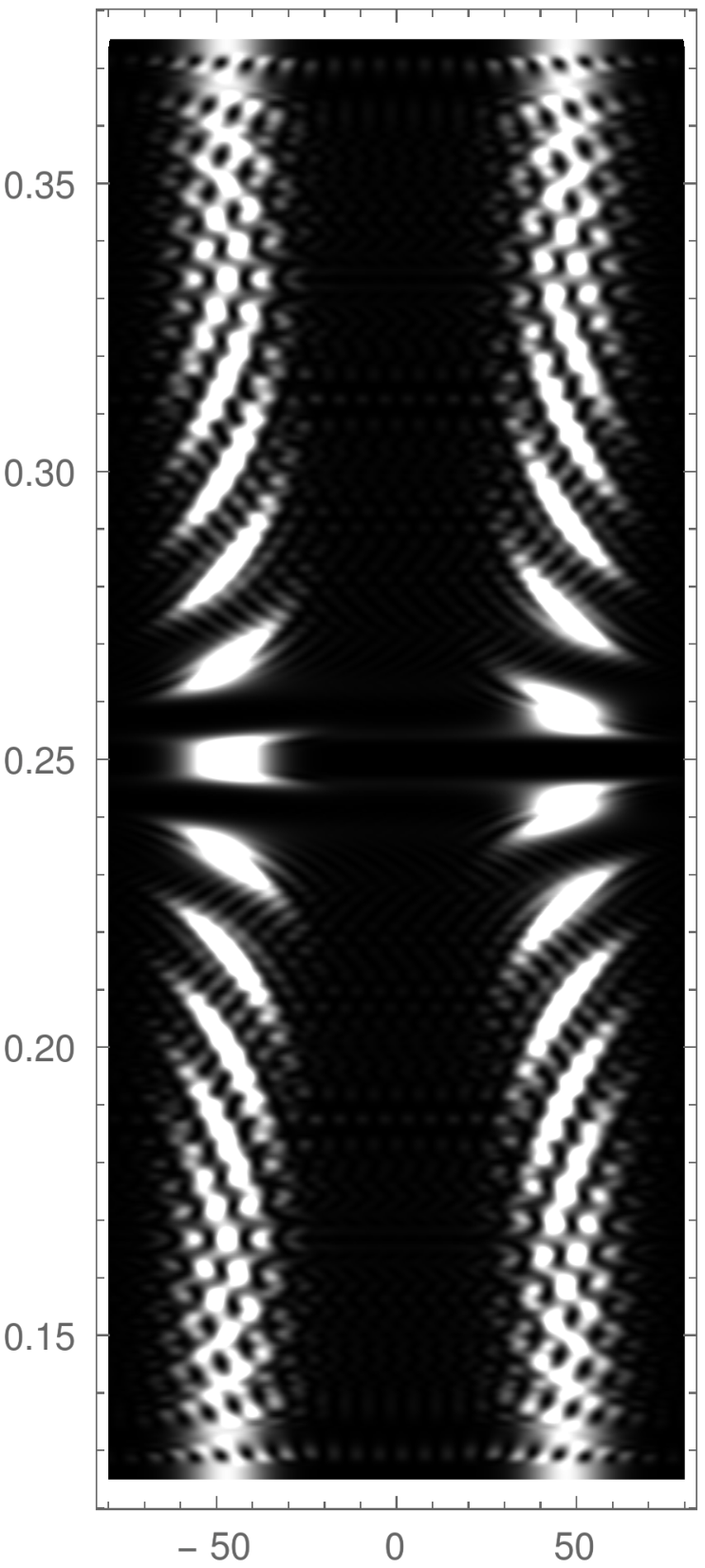}
\includegraphics[width=0.25\textwidth]{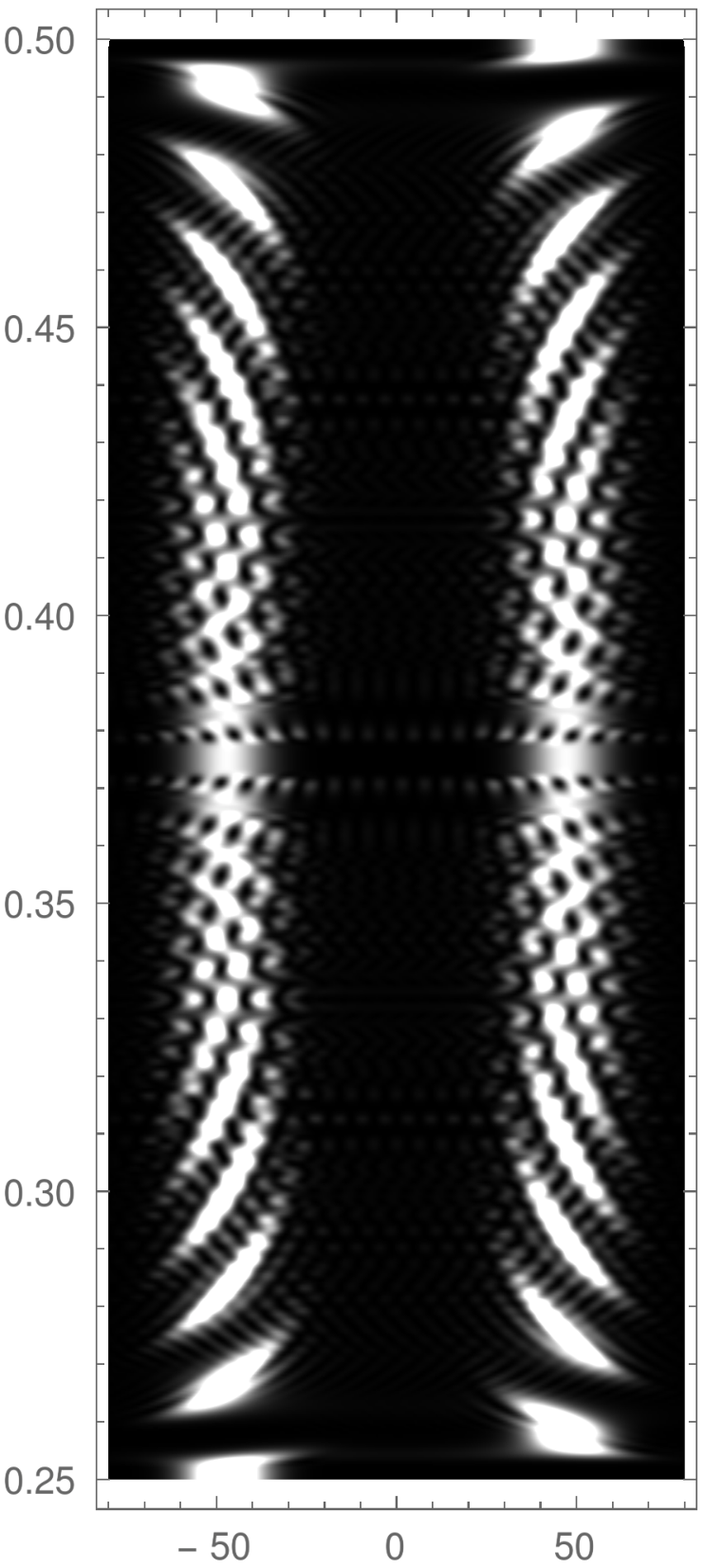}
\caption{Time evolution of momentum space probability density $\rho(p,t)=\vert\phi(p,t)\vert^{2}$ for an initial Gaussian wave packer with $p_{0}=15\pi$, highlighting different fractional revivals in a time window equal to $T_{rev}/2$.} \label{qcpfrac}
\end{figure}
In Fig. (\ref{qcxhalf}), the time evolution of position space probability density $\rho(x,t)$ for a Gaussian wave packer with initial momentum $p_{0}=30\pi$ is plotted for time $t=T_{rev}/2$ which explains the formation of quantum carpet in position space. It is obvious from the plot (left sub-plot) that a well-localized single peaked probability density of the wave packet evolves quasi-classically during its early time evolution and splits in to multiple sub-peaks after successive bounces with the walls of the deep square well where a self-interference of the wave packet takes place. These multiple sub-peaks then evolve in time with their own phases and undergo a series of constructive and destructive interferes (middle sub-plot). This results in the formation of regular arrangement of maximum probability regions$-$ bright fringes known as ridges $-$ and minimum probability regions $-$ dark fringes known as canals $-$ (right sub-plot).
During this course of time evolution, fractional copies of the original wave packet appear at times $t=T_{rev}p/q$ with $p,q$ are mutually prime numbers as shown in Fig. (\ref{qcxfrac}). In these plots, fractional revivals of the order $T_{rev}/10,T_{rev}/8,T_{rev}/6,T_{rev}/5,T_{rev}/4$, and their mirror fractions in the next quarter, can easily be identified.

\begin{figure}
\centering
\includegraphics[width=0.22\textwidth]{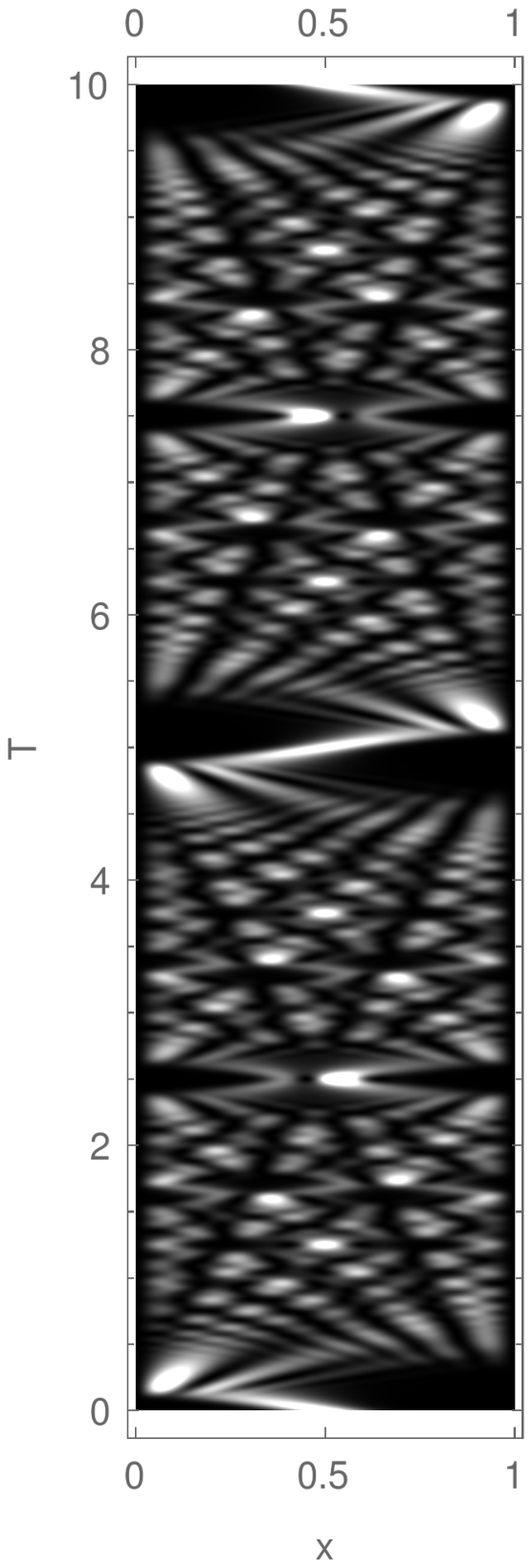}
\includegraphics[width=0.22\textwidth]{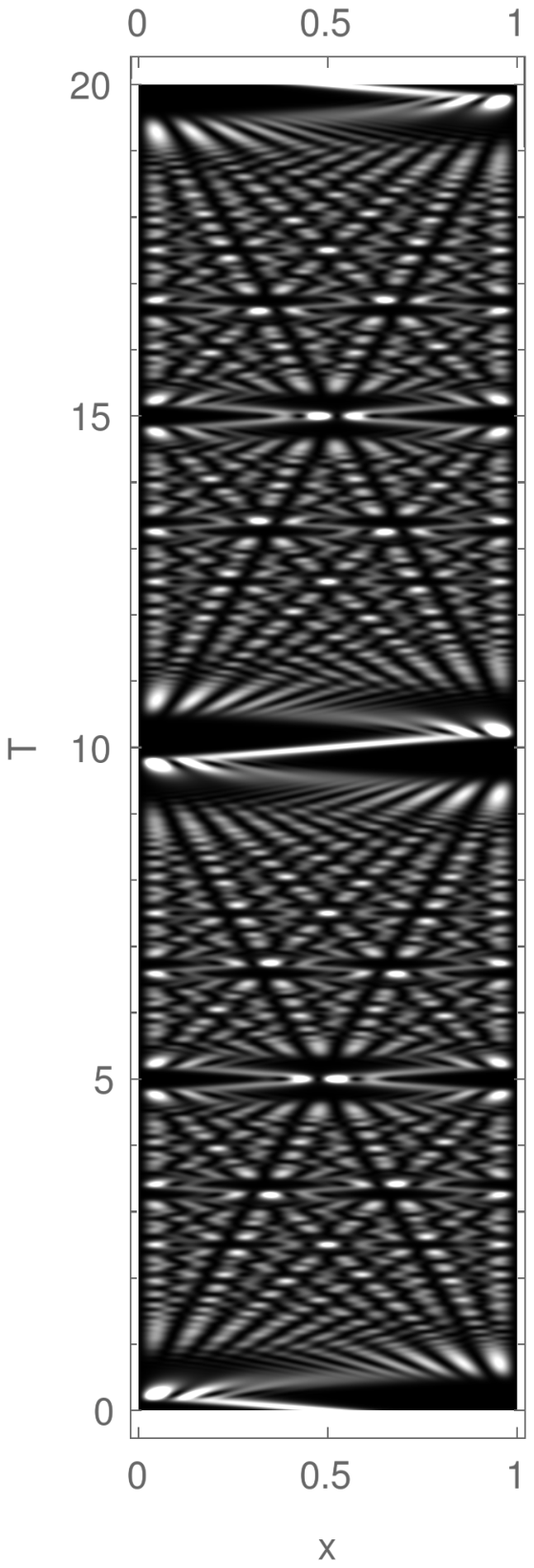}
\includegraphics[width=0.22\textwidth]{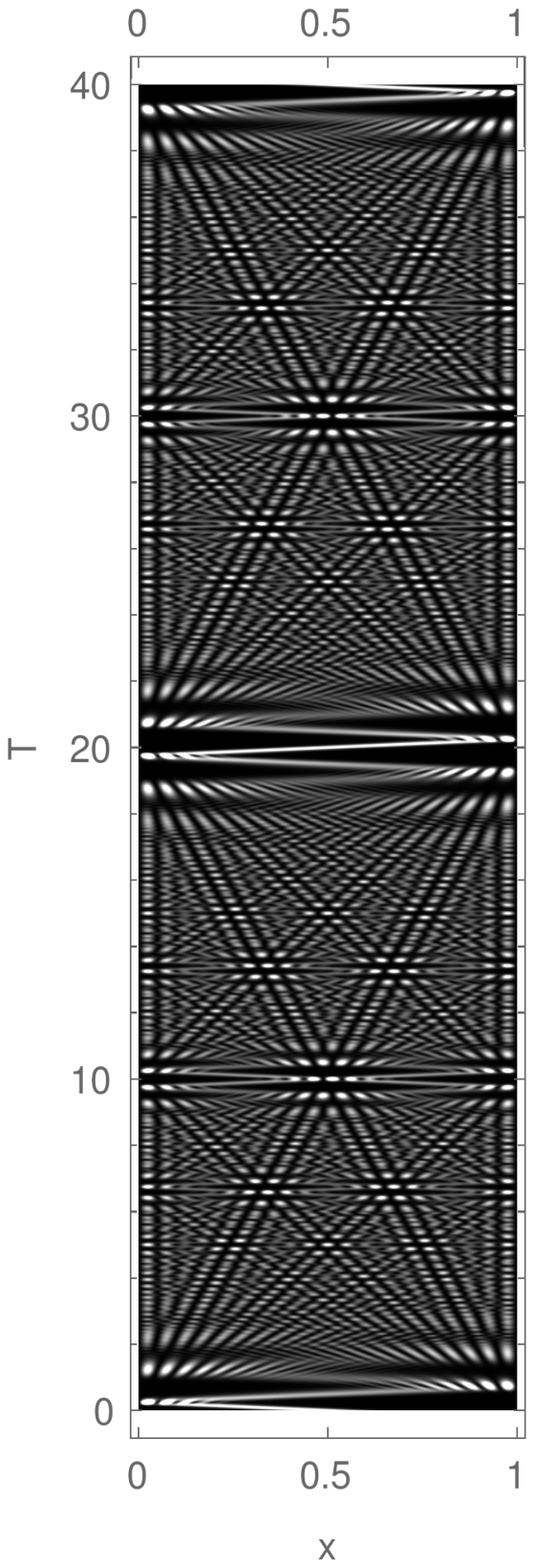}
\includegraphics[width=0.22\textwidth]{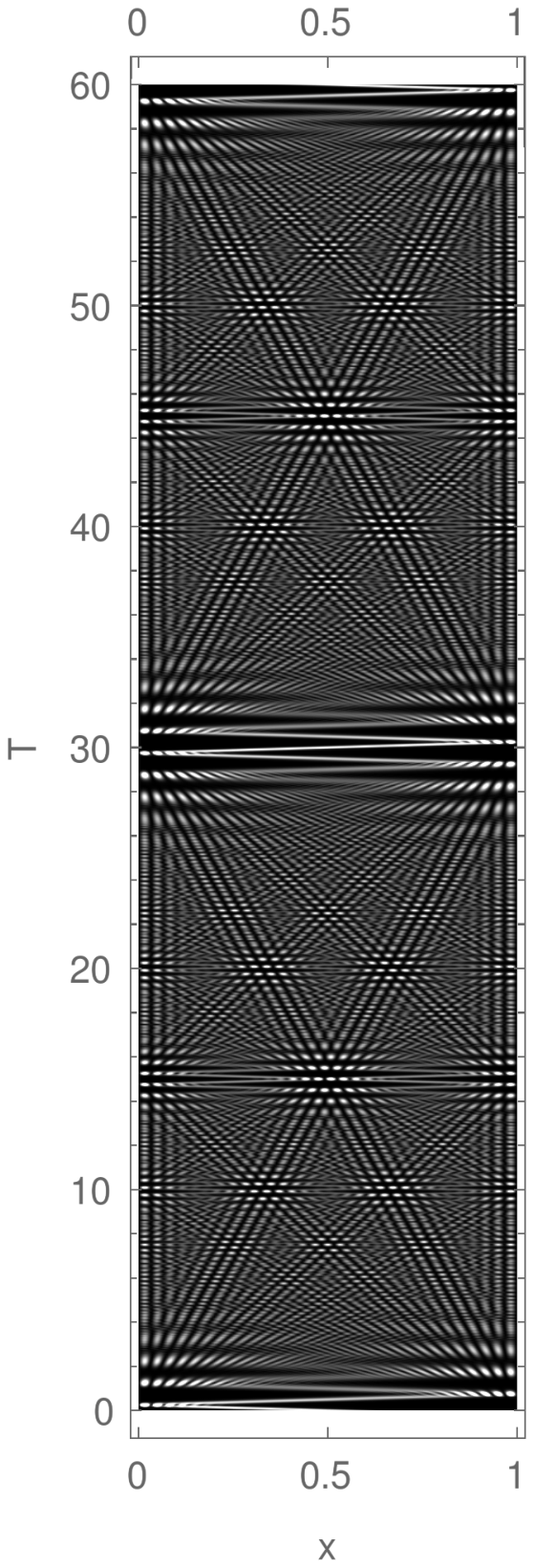}
\caption{Time evolution of $\rho(x,t)=\vert\psi(x,t)\vert^{2}$ for the same wave packer as in Figure (\ref{qcxhalf}) with different values of $p_{0}=5\pi,10\pi,20\pi,30\pi$ from left to right, respectively. A time period equal to one full quantum revival, $T_{rev}$ is plotted. } \label{qcxfull}
\end{figure}
\begin{figure}
\centering
\includegraphics[width=0.22\textwidth]{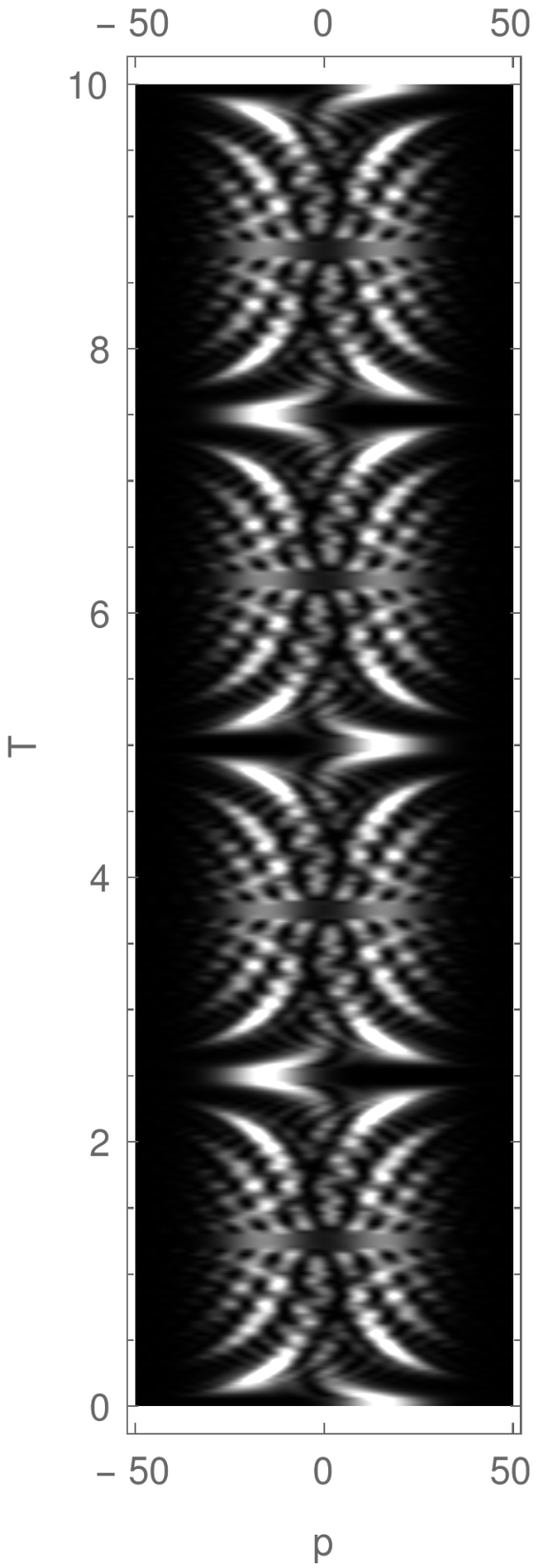}
\includegraphics[width=0.22\textwidth]{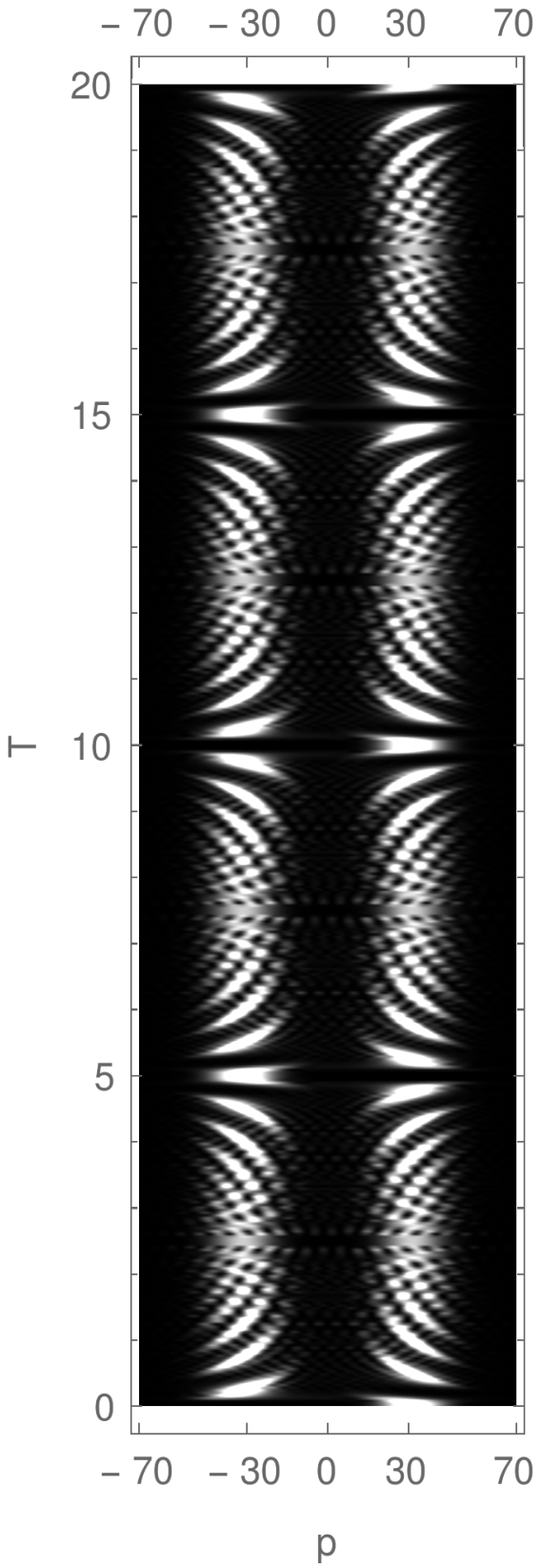}
\includegraphics[width=0.22\textwidth]{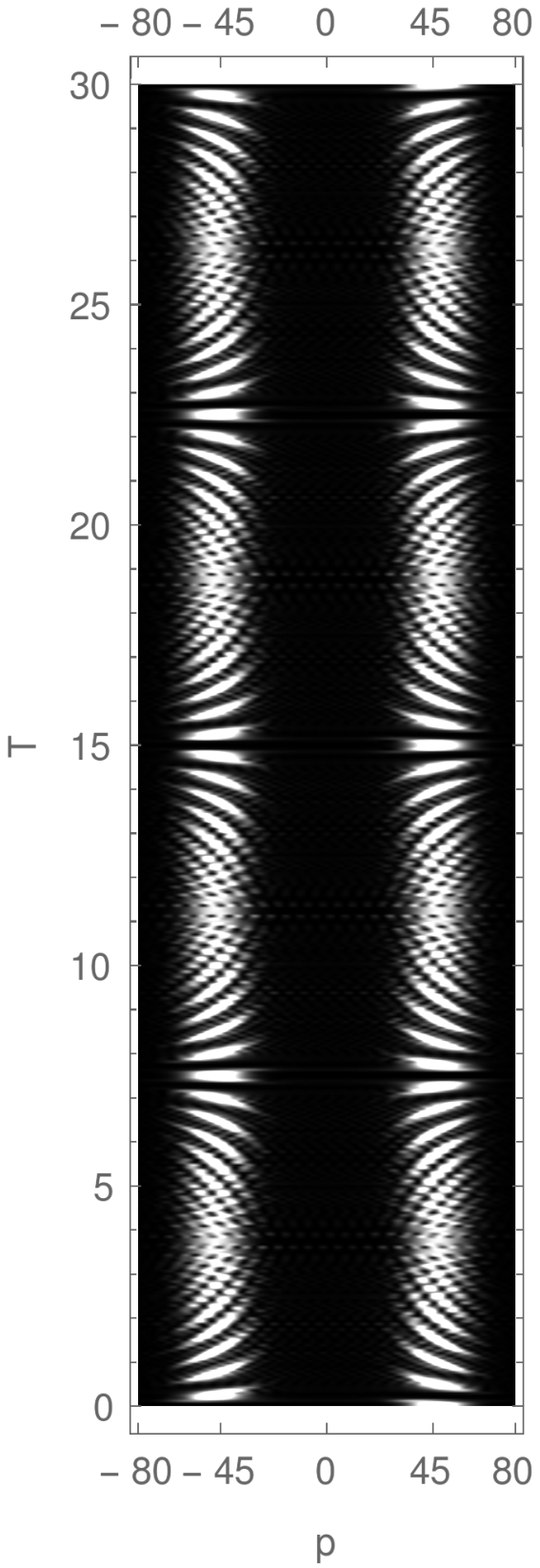}
\includegraphics[width=0.22\textwidth]{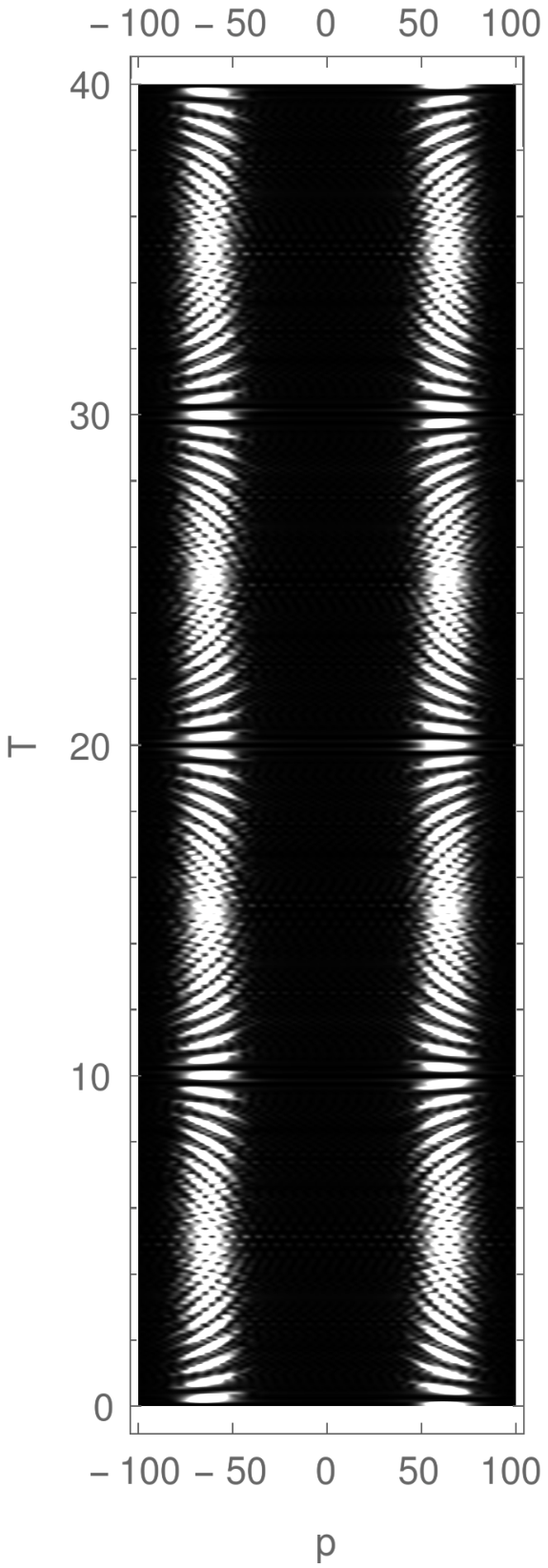}
\caption{Time evolution of $\rho(p,t)=\vert\phi(p,t)\vert^{2}$ for the same wave packer as in Figure (\ref{qcpfrac}) with different values of $p_{0}=5\pi,10\pi,15\pi,20\pi$ from left to right, respectively. A time period equal to one full quantum revival, $T_{rev}$ is plotted.} \label{qcpfull}
\end{figure}
The time evolution of momentum space probability density $\rho(p,t)=\vert\phi(p,t)\vert^{2}$ of the same wave packet, but with $p_{0}=15\pi$, is plotted in Fig. (\ref{qcpfrac}) for a time window of $T_{rev}/2$. It is seen that the momentum space carpets mimic different designs as compared to the position space carpets. These carpets also reflect the information of major significant fractional revivals. For the sake of completeness, we present a detailed gallery of quantum carpets, shown in Fig. (\ref{qcxfull}) and Fig. (\ref{qcpfull}), woven by position space and momentum space wave packets, respectively, excited at different mean quantum numbers $n_{0}$. It is obvious from these plots that the design of the carpets reflects more fine structure as the value of $p_{0}$ is increased and therefore the structure of fractional revivals is more prominent. 

\section{Summary and Conclusions}\label{sum}
In this work, dynamical recurrences of quantum wave packets have been discussed and the structure of quantum revivals and fractional revivals have been analyzed by means of temporal evolution of autocorrelation and wave packet probability densities, both, in position and momentum spaces. The temporal evolution of wave packet probability density, either in position or momentum space, undergoes a sequence of self-interference which leads to the formation of a regular interference pattern, namely, quantum carpets. We present a large gallery of quantum carpets woven by a Gaussian wave packet both in position and momentum spaces. It has been shown that the designs of these carpets reflect the self-similar structures after regular intervals of space-time and momentum-time which leads us to identify the structure of the quantum revivals and fractional revivals.
  
Due to its analytic simplicity and availability of its exact solutions, the model of particle-in-box  has extensively been studied in quantum mechanics which helped to understand a lot of physical phenomena of great importance, for instance, quantum information processing \cite{si08}, dynamical behavior of Fermi-Ulam accelerator \cite{ta}, temporal stability of generalized coherent states\cite{gk,gcs1}. This model is extremely helpful to understand the dynamical recurrences of the wave packet due to $n^{2}$ dependence of its energy spectrum which reflects a nice symmetry in structure of fractional revivals. Our analysis of identifying wave packet fractional revivals by means of quantum carpets can be extended for other systems of more practical interest.



\begin{thebibliography}{99}

%
%

\bibitem{jpb1} G. Alber, H. Ritsch and P. Zoller, {\it Phys. Rev. A}, {\bf 34}, 1058 (1986).

\bibitem{jpb2} J. Parker and C. R. Stroud Jr., {\it Phys. Rev. Lett.}, {\bf 56}, 716 (1986).

\bibitem{jpb3} J. A. Yeazell, M. Mallalieu and C.R. Stroud Jr., {\it Phys. Rev. Lett.}, {\bf 64}, 2007 (1990).

\bibitem{jpb17} R. W. Robinett,  {\it Phys. Rep.}, {\bf 392}, 1 (2004).

\bibitem{gcs1}  S. Iqbal, P. Rivi\'{e}re and F. Saif, \textit{Int. J. Theor. Phys.}, \textbf{49}, 2540 (2010).
\bibitem{gcs2}  S. Iqbal and F. Saif, \textit{J. Russ. Laser Res.}, \textbf{34}, 77 (2013).
\bibitem{gcs3}  N. Amir and S. Iqbal, \textit{J. Math. Phys.}, \textbf{56}, 062108 (2015).
               
\bibitem{si2006} S. Iqbal, Qurat-ul-Ann and F. Saif, \textit{Phys. Lett. A}, \textbf{356}, 231 (2006);\\
                 F. Saif, \textit{Phys. Rep.}, \textbf{419}, 207 (2005);\\
                 M. Ayub and F. Saif, \textit{Phys. Rev. A}, \textbf{85}, 023634 (2012).

\bibitem{jpb9} P. Rivi\'{e}re, S. Iqbal, and J-M Rost,  \textit{ J. Phys. B: At. Mol. Opt. Phys.}, \textbf{47}, 124039 (2014);\\
               P. Rivi\'{e}re, S. Iqbal, and J-M Rost, \textit{ J. Phys.: Conf. Ser.}, \textbf{194}, 022015 (2009);\\
               H. Katsuki, H. Chiba, C. Meier, et al., {\it Phys. Chem. Chem. Phys.}, {\bf 12} 5189 (2010).

\bibitem{jpb9a} J. Keeling and V. Gurarie, {\it Phys. Rev. Lett.}, {\bf 101},  033001 (2008).

\bibitem{jpb10} I. L. Garanovich, S. Longhi, A. A. Sukhorukov et al., {\it Phys. Rep.}, {\bf 518}, 1  (2012).

\bibitem{jpb11} V. Krueckl and T. Kramer, {\it New J. Phys.}, {\bf 11}, 093010  (2009);\\
                R. Ali, and F. Saif, \textit{Mater. Res. Express}, \textbf{2},  095602 (2015).


\bibitem{jpb13} C. O. Reinhold, S. Yoshida, J. Burgd\"{o}rfer, et al., {\it J. Phys. B: At. Mol. Opt. Phys.},
{\bf 42}, 091003 (2009).

\bibitem{jpb14} Z. D. Gaeta and C. R. Stroud Jr.,{\it Phys. Rev. A}, {\bf 42}, 6308  (1990).

\bibitem{jpb15} C. Leichtle, I. S. Averbukh and W. P. Schleich, {\it Phys. Rev. A}, {\bf 54}, 5299  (1996).

\bibitem{jpb16} D. L. Aronstein and C. R. Stroud Jr., {\it Laser Phys.}, {\bf 15}, 1496 (2005).


\bibitem{jpb18} I. S. Averbukh and N. F. Perelman, {\it Phys. Lett. A}, {\bf 139}, 449 (1989).

\bibitem{jpb19} R. Veilande and I. Bersons, {\it J. Phys. B: At. Mol. Opt. Phys.}, {\bf 40}, 2111 (2007).

\bibitem{jpb20} J. A. Yeazell and C. R. Stroud Jr., {\it Phys. Rev. A}, {\bf 43}, 5153 (1991).

\bibitem{jpb21} G. D. Valle, M. Savoini, M. Ornigotti, et al., {\it Phys. Rev. Lett.}, {\bf 102}, 180402 (2009).

\bibitem{jpb22} S. Ghosh and J. Banerji, {\it J. Phys. B: At. Mol. Opt. Phys.}, {\bf 40}, 3545 (2007).

\bibitem{jpb23} A. Schubert, K. Renziehausen and V. Engel, {\it Phys. Rev. A}, {\bf 82}, 013419 (2010).

\bibitem{jpb24} W. Merkel, S. I. Averbukh, B. Girard, et al.,  {\it Fortschr. Phys.}, {\bf 54}, 856 (2006).

\bibitem{jpb25} E. Romera and F. de los Santos,{\it Phys. Rev. Lett.}, {\bf 99}, 263601 (2007);\\
                E. Romera and F. de los Santos, {\it Phys. Rev. A}, {\bf 78}, 013837  (2008);\\
                T. Abbas and F. Saif, \textit{Int. J. Theor. Phys.}, \textbf{53}, 1961  (2014).

\bibitem{jpb26} V. Ayvazyan N. Baboi, J. B\"{a}hr, et al, {\it Eur. Phys. J. D}, {\bf 37}, 297  (2006).
\bibitem{qc1} F. Grossmann, J-M Rost and W. P. Schleich, {\it J. Phys. A: Math. Gen.}, {\bf 30},L277 (1997).
\bibitem{qc2}  I. Marzoli, F. Saif, I. Bialynicki-Birula, et al., {\it Acta Physica Slovaca}, {\bf 48}, 323 (1998).
\bibitem{qc3} O. M. Friesch, I. Marzoli and W. P. Schleich, {\it New J. Phys.}, {\bf 2}, 4 (2000).
\bibitem{qc4} P. Kazemi, Chaturvedi, I. Marzoli, R. F. O'Connell et al., {\it New J. Phys.}, {\bf 15}, 013052 (2013).
\bibitem{qc5} F. Saif, Las Phys., \textbf{22}, 1874 (2012).
\bibitem{qc6} S. V. Prants, V. O. Vitkovsky, {\it J. Russ. Laser Res.}, {\bf 31}, 201  (2010).
\bibitem{si08} S. Iqbal and F. Saif, {\it J. Russ. Laser Res.}, \textbf{29}, 466  (2008).
\bibitem{ta} T. Abbas and F. Saif, {\it J. Russ. Laser Res.}, \textbf{33}  448 (2012).
\bibitem{gk} J. P. Antoine, J. P. Gazeau, P. Monceau et al., {\it J. Math. Phys.},
{\bf 42}, 2349 (2001).

\end{thebibliography}
\end{document}